\documentclass[onecolumn,aps,pra,superscriptaddress,amsmath,amssym,floatfix,longbibliography,notitlepage,amsbook]{revtex4}
\usepackage[colorlinks=true,linkcolor=blue,citecolor=blue,filecolor=green,urlcolor=blue]{hyperref}
\usepackage{amsmath}
\usepackage{graphicx}
\usepackage{algorithm}
\usepackage{algpseudocode}
\usepackage{physics}
\newcommand\be{\begin{equation}}
\newcommand\ee{\end{equation}}
\newcommand\la{\langle}
\newcommand{\ra}{\rangle}

\begin{document}

\title{Shot Noise Limited Triangulation}

\author{John C. Howell}
\affiliation{Institute for Quantum Studies, Chapman University, Orange, CA 92866, USA}
\affiliation{Racah Institute of Physics, The Hebrew University of Jerusalem, Jerusalem, Israel, 91904}
\author{Andrew N. Jordan}
\affiliation{Institute for Quantum Studies, Chapman University, Orange, CA 92866, USA}
\affiliation{The Kennedy Chair in Physics, Chapman University, Orange, CA 92866, USA}

\email{johhowell@chapman.edu}

\begin{abstract}
We design a system-level architecture for approaching the shot noise limit for passive triangulation of a quasi-monochromatic point source. Our emphasis is not in the novelty of the basic physics, but that existing systems lose fundamental information in the measurement pipeline. We preserve that information through maintaining analog signals combined with common-mode noise rejection in the layers of signal processing.  We review the Cram\'er-Rao bound of angle sensing as applied to the field of triangulation. Using a monolithic camera/balanced detector system and a doubly-layered analog voltage differential system, we experimentally achieve nanometer-scale depth precision at 1.42 meter standoff with a baseline of only 10 centimeters. While still roughly two orders of magnitude above the shot noise limit, the results represent several orders of magnitude improvement over current camera-only or other position-sensitive detector systems. The system can be further improved with vibration and turbulence mitigation.    
            
\end{abstract}

\maketitle
\section{Introduction}
Triangulation utilizes the angular disparity of an object as viewed by two laterally-displaced observations \cite{hartley1997triangulation,lindstrom2010triangulation,hendeby2006triangulation,lee2019angular,henry2023absolute} to calculate the relative 3D position of the object with respect to the viewer.  Eye-brain \cite{ponce2008stereopsis} and machine vision systems \cite{szeliski2022computer} can achieve sub-mm position precision at meter standoffs using this technique. Owing to its great importance in a wide range of applications \cite{farina1999bearing,dogancay2005total,biswas2012depth,Paulus2014,masoumian2022monocular,dong2022towards}, much effort is being devoted to artificially intelligent stereo 3D reconstruction \cite{Laga_2022}.  Importantly, while the precision of the lateral coordinates of an object in triangulation scale linearly with the distance of the object, the depth precision scales quadratically. This quadratic behavior can make triangulation a poor depth estimator when the object distance greatly exceeds the baseline (the distance between detectors, e.g., the eyes).  Here, we discuss and demonstrate dramatic improvements in depth estimation of passive point sources over current systems \cite{Laga_2022,meng2022six,cheng20233d, zeng1999two,liu2004development} by pushing our system towards its fundamental limits.  We outline the fundamental underlying theory that limits triangulated point source localization based on well-known angle sensing limits \cite{treps2003quantum}.  We then experimentally demonstrate a camera-directed balanced detector system to achieve nanometer depth precision at meter standoff distances with only a ten centimeter baseline. We note that the parallax sensitivities achieved here are comparable to those for the Gaia star surveyor \cite{prusti2016gaia}, albeit in a very different context.    

Depth estimation is of critical importance in many applications including: robotics, navigation, autonomous vehicles, augmented reality, aerial mapping, medical imaging, quality control, vision science etc..\cite{farina1999bearing,dogancay2005total,biswas2012depth,Paulus2014,masoumian2022monocular,dong2022towards}.  Passive, remote, depth estimation, primarily through triangulation, has a long and rich history dating back millenia \cite{henry2023absolute} and has been a critical tool in navigation \cite{kaplan2011angles} and surveying \cite{lindgren199019} for several centuries. In its modern use, triangulation is the backbone for depth estimation in passive stereo imaging driving major research efforts in algorithmic and artificially intelligent reconstruction of 3D scenes \cite{Laga_2022}. Unfortunately, camera-based systems have typically been limited to precision limits greater than 10's of microns (for a review of various stereo 3D reconstruction, see \cite{Laga_2022}). We note that this work is distinct from active systems, such as Lidar, where phase retrieval can achieve nanometer-scale distance precision with ease.   

There is a rich history on triangulation limits.  Dorsch, Hausler and Herrmann showed that triangulation through laser-illuminated targets was limited by speckle to between 1 and 10 microns. Hartley and Sturm \cite{hartley1997triangulation} and updated by Lindstrom \cite{lindstrom2010triangulation} derived practical, but not fundamental, limits based on algorithmic and geometric considerations.  Clark and Ivekovic \cite{clark2010cramer} outline Cram\'er-Rao bounds for 3D state estimation from stereo cameras based on camera properties and estimation techniques.  Recent efforts have been devoted to overcoming speckle limits \cite{willomitzer2019state}.  Here, we take a different approach. Our fundamental limits are not based on technical limitations or speckle statistics of coherent illumination, but on the photon statistics of the source.  

While dual-camera systems can generate quasi-continuous 3D point clouds of scenes, more specialized systems measure single point sources with the intent of achieving  higher accuracy. In this category, methods based on position sensitive detectors utilizing the quadrant detector and lateral effect photodetector, have shown excellent results and promise.  For example, the lateral effect pincushion position sensitive detector binoculars have achieved sub-millimeter precision accuracy \cite{meng2022six,cheng20233d}, but have faced challenges with the nonlinearity of the detectors.  Active triangulation using laser illumination of targets such as mirrors or gratings have shown millimeter down to micron accuracy \cite{Paulus2014,zeng1999two,liu2004development}.  A proposal for a passive triangulation system was made in \cite{olyaee20123}, but no experiment was performed or theoretical limits outlined.  It should also be noted that the most accurate triangulation experiment performed to date was the Gaia spacecraft star mapper \cite{prusti2016gaia}, which used a combination of differential background referencing, time-delayed integration and precision timing to obtain 100 prad parallax accuracy to achieve unprecedented intra-galactic stellar-distance measurements, the angular precision we achieve in this work.  

Here, we discuss theoretically and experimentally demonstrate passive dual-balanced-detector stereo depth estimation of point sources at the nanometer-sensitivity scale. We believe the novelty of this work lies in designing a system-level architecture, based on the analysis of the fundamental equations governing triangulation, that minimizes the information loss in the measurement pipeline. Secondly, while the Cram\'er-Rao bound for angle sensing is well known, we have not found its use in triangulation, even though this may seem like an obvious next step. Our Fisher-information preserving system represents a departure from existing systems which treat measurement noise as external parameter rather than as fundamental for the reconstruction\cite{hartley1997triangulation,lindstrom2010triangulation}.

The paper is organized as follows: In section \ref{Theory}, we derive the fundamental theory of point source location via triangulation.  In section \ref{Experiment}, we discuss the experimental system we designed to achieve high precision depth estimation and consider the results in section \ref{Results}.  We finish with some discussions of advantages and limitations in section \ref{Discussion} and wrap up with conclusions in section \ref{Conclusions}. We have also added an appendix which discusses the 3D generalizations of the triangulation.  

\section{Theory: Triangulated Point Source Localization} \label{Theory}
  
\subsection{Basics of Triangulation}
\begin{figure}[ht]
    \includegraphics[width=.48\textwidth]{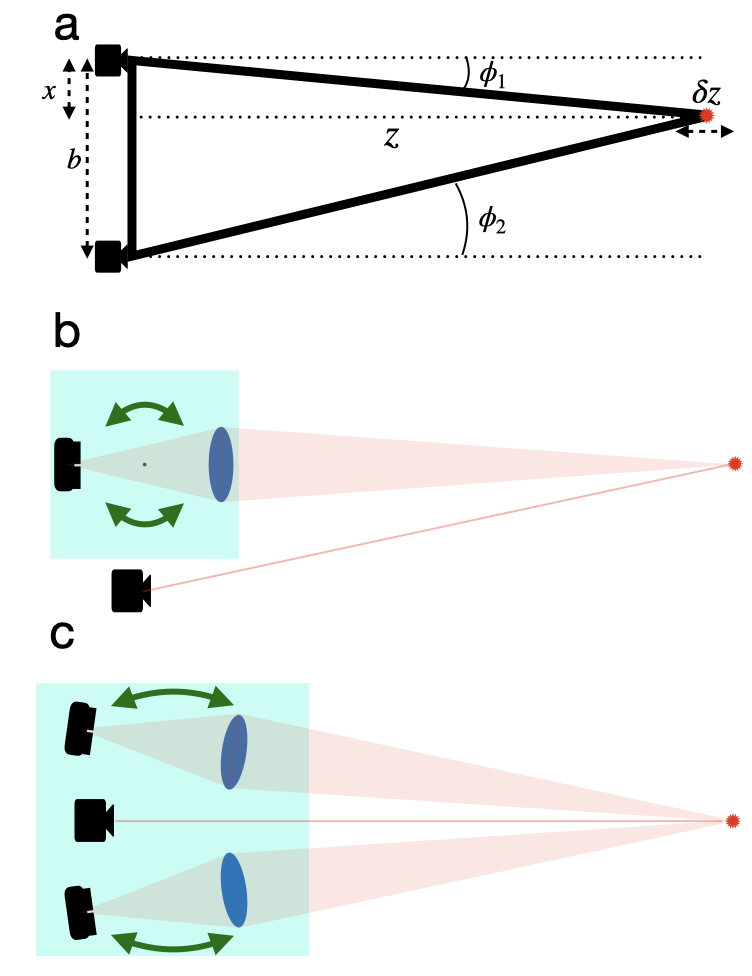}
    \caption{Stereoscopic depth estimation experiments.  a) shows the standard technique of using two high resolution cameras to assess the position of a point source at $(x,z)$. A single camera in a) is replaced by the two-element sensor in b). In b) the camera gives a rough approximation of the position of a point source.  Then, a movable lens/quad-segmented photodiode system orients itself toward the point source at nearly optimal sensitivity. If the movements can be encoded at sub-nRad accuracy, the system achieves nearly the shot noise limit of sensitivity. Panel c) shows a high-accuracy, low-cost monolithic system used in this paper.  A single camera is used for obtaining transverse coordinates and dual balanced detectors make high-accuracy depth measurements.  }
    \label{fig:Schematic}    
\end{figure}

The principles of triangulation are well known (see for example \cite{Blostein1987,hartley1997triangulation,lindstrom2010triangulation,hendeby2006triangulation,lee2019angular,henry2023absolute}). Here, we give a short review of the basic physics and mathematics.  An experimental schematic of standard stereo depth estimation of a point source is shown in Fig. \ref{fig:Schematic} a).  Two angle-sensitive sensors (e.g., cameras, quadrant detectors, position sensitive detectors) laterally separated by a baseline distance $b$ are used to remotely determine the three-dimensional position of a point source with depth uncertainty $\delta z$.  For single parameter estimation, namely the uncertainty in depth, we constrain the problem to the plane which contains the two sensors and the point source, thus requiring us to only find the $(x,z)$ coordinate of the source.  While we treat the full 3D problem later, here we will focus on the depth, since it is the primary difficulty of passive point source localization. 

Straightforward calculations show that the depth is given by 
\begin{equation}
    z=\frac{b}{\tan(\phi_1)+\tan(\phi_2)}.
    \label{Z}
\end{equation}
Consider the simplifying situation in which $\phi=\phi_1=\phi_2$ resulting in
\begin{equation}
    z=\frac{b }{2 \tan(\phi)}.   
\end{equation}
To determine the depth uncertainty based on the uncertainty in angle and baseline, we take the differential and rewrite it in terms of $z$ to find
\begin{equation}
    \delta z=\delta b\frac{z}{b}-\delta \phi\left(\frac{b}{2} +\frac{2z^2}{b}\right).
    \label{deltaz}
\end{equation}
This equation tells us the domains in which various uncertainties dominate.  For example, in the regime where $b$ is known with a high accuracy and $z\gg b$, the uncertainty in depth is given by angular disparity and grows quadratically with depth. For multi-parameter estimation, namely when both angles are allowed to vary, we find the result
\begin{equation}
    \delta z = -\frac{x^2 + z^2}{b} \delta \phi_1 -  \frac{(b-x)^2 + z^2}{b}  \delta \phi_2 +  \frac{z}{b} \delta b.
\end{equation}
In the Appendix, we share the full three-dimensional equations for point source localization.  

\subsection{Angle of Arrival Uncertainty Bounds}
We now consider the fundamental angular uncertainty $\delta \phi$ relative to the optical axis of a lens/detector system \cite{treps2003quantum,dixon2009ultrasensitive}.  We assume a sensor is placed in the focal plane of a circular lens with focal length $f$ and diameter $D$. This brings the incoming light to a focus at the sensor.  The point source can be coherent or incoherent.  Since we are imaging a point onto the sensor, the intensity profile of the point source on the sensor will be given by the Airy disk or the intensity point spread function (PSF) $f_c(\rho)$ of a circular aperture:  
\begin{equation}
    f_c(\rho)=I_0\left(2\frac{J_1(kD\rho/2f)}{kD\rho/2f}\right),
\end{equation}
where, $J_1$ is the first order Bessel function of the first kind, $\rho$ is the radial distance from the center of the disk, $k$ is the central wavenumber (we assume a narrow band source) and $I_0$ is the intensity at the center.  However, we use a Gaussian approximation $f_G(x,y)$ of the PSF given by 
\begin{equation}
    f_G(x,y)=I_0 \exp\left(\frac{-x^2-y^2}{2\sigma^2}\right),
\end{equation}
where $\sigma\approx\lambda f/D$.  We do this for two reasons.  First, the PSF is separable in $x$ and $y$ allowing us treat this as a one-dimensional problem.  Second, as we will show below, the Fisher information is simple to compute for Gaussian  functions, namely  $N/\sigma^2$ for $N$ independent measurements (photons in this case).  

The Fisher information for a parameter $g$ of a continuous probability distribution $p_g(x)$ is defined as 
\be
I_g = \int dx (\partial_g \ln p_g(x))^2 p_g(x).
\ee
By geometric optics, a deflection angle $\phi$ results in a shift of the centroid of the Gaussian beam by an amount $f \phi$, so the relevant probability distribution function is given by $p_\phi(x, y) = (x - f \phi, y)$.
Our interest is to compute the Fisher information for determining the mean angle $\phi$ relative to the optical axis and not the mean (centroid), $d = f \phi$, of the PSF distribution, so we can convert the Fisher information as $I_\phi = f^2 I_d$.  For $N$ independent measurements the distribution factorized for uncorrelated measurement, we find the Fisher information is straight forwardly calculated to be \cite{cramer1999mathematical}
\begin{equation}
    I_d=\frac{N}{\sigma^2}, \quad I_\phi =
\frac{N}{(\lambda/D)^2}.
\end{equation}
The Cram\'er-Rao bound states that the best case variance of any unbiased estimator $\hat g$ for large data sets is bounded by the inverse Fisher information,
\be
\la {\hat g}^2\ra \ge I^{-1}.
\ee
Applied to our case of angle estimation, these principles lead to a Cram\'er-Rao lower bound of
\begin{equation}
    \delta \phi \gtrsim\frac{\lambda}{\sqrt{N}D},
    \label{CRB}
\end{equation}
where $\delta \phi$ is an unbiased uncertainty (standard deviation) of the mean angle. We note that this implies a single wavelength $\lambda$ and not a broadband source.  However, we will assume that the source is narrowband ($\Delta \lambda \ll \lambda$ e.g., single-color LED), but not necessarily quasi-monochromatic (e.g., laser), making this bound a reasonable approximation. Putting this all together via Eq.~(\ref{deltaz}), we obtain
\begin{equation}
    \delta z \geq\frac{\lambda}{\sqrt{N}D}\frac{2z^2}{b}.
\end{equation}

To put some numbers to this bound, a 1 $\mu W$ measured signal of 500 nm light in a 25 mm aperture would yield an angular lower bound of approximately 10 pRad Hz$^{-1/2}$.  Assuming a standoff distance of 1.42 m and a baseline distance is exactly $b=10$ cm, the depth uncertainty is then calculated to be approximately 500 pm Hz$^{-1/2}$, and increases to 13 nm Hz$^{-1/2}$ at 10 m standoff.   

\subsection{Triangulation Cram\'er-Rao Bounds}
An obvious follow up question is why existing stereo systems have not achieved this remarkable precision.  There are a few reasons.  Fundamentally, digitization prior to differential measurements leads to digitization noise, discretization of the signal and readout noise.  Consider a simple example: with the use of nearest-neighbor digitized gray-scale statistics, most commercial systems are often technically limited to approximately 0.1 pixel precision.  Assuming a 2 micron pixel and a 100mm focal length lens, the best angular resolution is approximately 2 microradians. Using the same dimensions described above for our system, that leads to a depth uncertainty of 73 microns at 1.42 meters.  To be competitive with our system would require 0.00001 pixel uncertainty in this scenario. These technical limitations effectively destroy Fisher information in the angular estimation processing stage. Since the depth estimation scales as the distanced squared, loss of angular Fisher information effectively gets amplified in the depth estimate.    

Balanced detection, on the other hand, can achieve far superior angular precision. Balanced detection firstly differences the analog voltages and then secondly digitizes the differential signal in contrast to most machine vision systems.  Differencing the analog output transimpedance voltages of a balanced detector allows for high common mode noise rejection.  As we will also show, a careful evaluation of the depth equation shows that we can further enhance the precision of the depth estimation with another differencing of analog signals between the balanced detectors.  Effectively, differencing the outputs of two balanced detectors allows for common-mode noise rejection between detectors.  For example, LED power fluctuations would be seen on both detectors.  This allows us to better approach the fundamental limits of precision depth estimation. As a note, the use of two balanced detectors does not improve the angular resolution, but utilizes the angular sensitivity of balanced detection for dual-detector triangulation.

Consider an experimental system with the ability to improve passive depth estimation by many orders of magnitude. One side of the modified point estimation experiment is shown in Fig. \ref{fig:Schematic} b).  In this figure, a single camera is replaced by a camera combined with a lens/quad segmented photodiode system able to operate at nearly the fundamental limit.  The camera obtains a rough estimation of the angle of the point source while the lens/photodiode system is on a movable platform that controls both the pitch and yaw of the lens/photodiode system.  By using accurate angular calibration combined with high-accuracy rotation encoding, it is possible to estimate the angle of arrival of light from the point source leading to hyper-precise depth estimation when combined with the other lens/photodiode system.   

The lens/detector system operates based on the idea that once a detector and lens are placed, they define an optical axis.  The detector is placed in the focal plane of the lens.  The lens, as we showed earlier, causes the point spread function of the source to be distributed across the focal plane. Here, we use quad detector as the photodetector, whose use as a precision angle sensor is well known \cite{treps2003quantum,dixon2009ultrasensitive}.  A quad detector uses the differential in measured voltage between its left and right and up and down segment combinations to determine the centroid of the point spread function. For example, with $N$ photons arriving from the source and a point spread function of width $\sigma$, a lateral shift of the point spread function by $\epsilon\ll \sigma$ causes a relative  shift of 
\begin{equation}
    \Delta N=\frac{\sqrt{2}\epsilon }{\sqrt{\pi}\sigma}N
\end{equation}
photons to be shifted from one side to the other.  Since the number of photons measured in each photodiode is linear in the number of total photons arriving, then dividing by the total number of photons on either side is
\begin{equation}
    \frac{\Delta N}{N}=\frac{\sqrt{2}\epsilon }{\sqrt{\pi}\sigma}=\frac{V_L-V_R}{V_L+V_R}\equiv B,
\end{equation}
where $B$ is defined as the normalized balanced detector signal.  Assuming we are still only in two dimensions and the axis of calibration is in the forward direction, we obtain $\phi$, namely:
\begin{equation}
    \phi=\alpha+\beta,
    \label{phi}
\end{equation}
where $\alpha$ is the yaw angle of the platform relative to axis of calibration in the forward direction and angle $\beta=\frac{\sqrt{\pi}\sigma B}{\sqrt{2 }f}$ is set by the remaining offset from the optical axis of the lens/quad detector system, where we have made the assumption that $\alpha \approx \phi$ so that $\epsilon\ll \sigma$.   

The uncertainty in the angle of arrival for this experimental configuration is set by the uncertainty in the number of photons measured in a time interval.  If we assume that the source obeys Poisson statistics, we can calculate the angular uncertainty.  In this work, we will assume that the variance of the number of photons $N$ arriving at the detector is $N$.  The uncertainty in the angle $\beta$ is then given by 
\begin{equation}
    \delta \beta=\frac{\sqrt{\pi}\sigma }{\sqrt{2N} f}=\sqrt{\frac{\pi}{2}}\frac{\lambda }{\sqrt{N} D},
\end{equation}
which can be seen to only be $\sqrt{\frac{\pi}{2}}$ larger than the Cram\'er-Rao lower bound (see Eqn. \ref{CRB}). Hence, the shot noise limit of a balanced detector system system is only slightly worse than the best possible estimate. 

\section{Experiment: Triangulated Depth Estimation} \label{Experiment}

While the Cram\'er-Rao bound gives a lower bound to the precision of a measurement, it does not specify the accuracy of a measurement, which is crucial for depth estimation. To achieve high accuracy, it is critical to know the value of the yaw angles ($\alpha_1$ and $\alpha_2$) for the two detectors in Eqn. \ref{phi} for Eqn. \ref{Z} of the platforms holding the balanced detectors.  Nanoradian accuracy optical encoders can achieve very high accuracy rotation stages for the balanced detectors.  However, we show a simple solution that does not require high-accuracy optical encoders, but only a simple calibration.  

We designed the experimental system as shown in Fig. \ref{fig:Schematic} c), utilizing a few key assumptions and simplifications.  First, we assume that the distance to the object greatly exceeds the baseline distance (i.e., $z\gg b$), which is true in most remote sensing contexts. In this small angle limit, we use $\tan(\phi)=\phi$.  Our calibration distance is $z=1.42$ m and the baseline distance $b=10$ cm.  As a note, one might argue that given the required precision $\alpha$ is still fairly large and thus the small angle approximation is not valid.  But, we point out that in most remote sensing applications, the baseline distance will be many orders of magnitude smaller than the object distance making the small angle approximation valid. Second, we used a single camera for gaining estimates of the transverse coordinates since they scale linearly with distance and not quadratically like depth estimates.  Third, we ensure accuracy of the yaw angles by building all of the components on a single monolithic platform and calibrating beforehand. The combination of these last two points removes the accuracy dependence of the platform in estimating the depth and puts the dependence on the initial calibration of the fixed yaw angles and balanced detection measurements. With this design, we obtain a valuable result, namely,
\begin{equation}
    z\approx\frac{b}{\alpha_1+\alpha_2+\beta_1+\beta_2}.
\end{equation}
While, we have considered the positive angle for $\phi_1$ to be clockwise relative to the optical axis and $\phi_2$ to be counter clockwise, the voltages both change in the same direction with respect to the optical axis.  This means that to measure $\beta_1+\beta_2$, we actually take an analog difference measurement of the differential signals from each detector (a second differential). This allows for further common mode noise rejection and amplification.  

\subsection{Experimental Setup}
We tested over a dozen different detectors during this experiment.  We built several in-house detectors based on the Hamamatsu two-element segmented photodiodes and used two New Focus 2901 Quadrant Detectors. The New Focus quadrant detectors were chosen because of their low-noise, low-bias, and adjustable transimpedance amplification.  The differential and sum signals are low-pass filtered and sent to a 24 bit, single-sided analog to digital converter (ADC).  However, we found that keeping the voltage amplitudes under 0.5 V allowed us to use the ADC as if it were double-sided and thus negating the need to bias the voltages.  The sampling rate of the system is 1 kHz.  We use a Texas Instruments INA118 instrumentation amplifier for the second differential voltage measurement between the two detectors.    

The detector system is built on a rotating platform consisting of a stepper motor which drives a worm gear rotation stage.  Each complete revolution of the stepper motor causes approximately 2 degrees of rotation of the stage allowing for sub micro-radian rotation control and accuracy.  A 12 Megapixel camera is centered on the stage between the two balanced detectors.  The camera feed is used to guide the rotation stage to the point sources where the balanced detectors are then able to measure the distance. The camera, stepper motors and analog to digital converter are all connected to a central computer used to control the system.  

For the light source, we used a small-profile LED with a measured signal power at the detector of approximately 1 $\mu W$. A surface mounted design LED was used because of its good illumination uniformity. The LED was placed on one of two mounts: a fixed mount with a piezoactuator for small precision translations or atop a stepper motor driven translation stage with micron step sizes.  The fixed mount and the translation stage were both on a non-floating optical table without any further temperature, vibration or turbulence isolation.   

\section{Experimental Results and Analysis}
\label{Results}
\begin{figure}[H]
    \centering
    \includegraphics[width=0.45\linewidth]{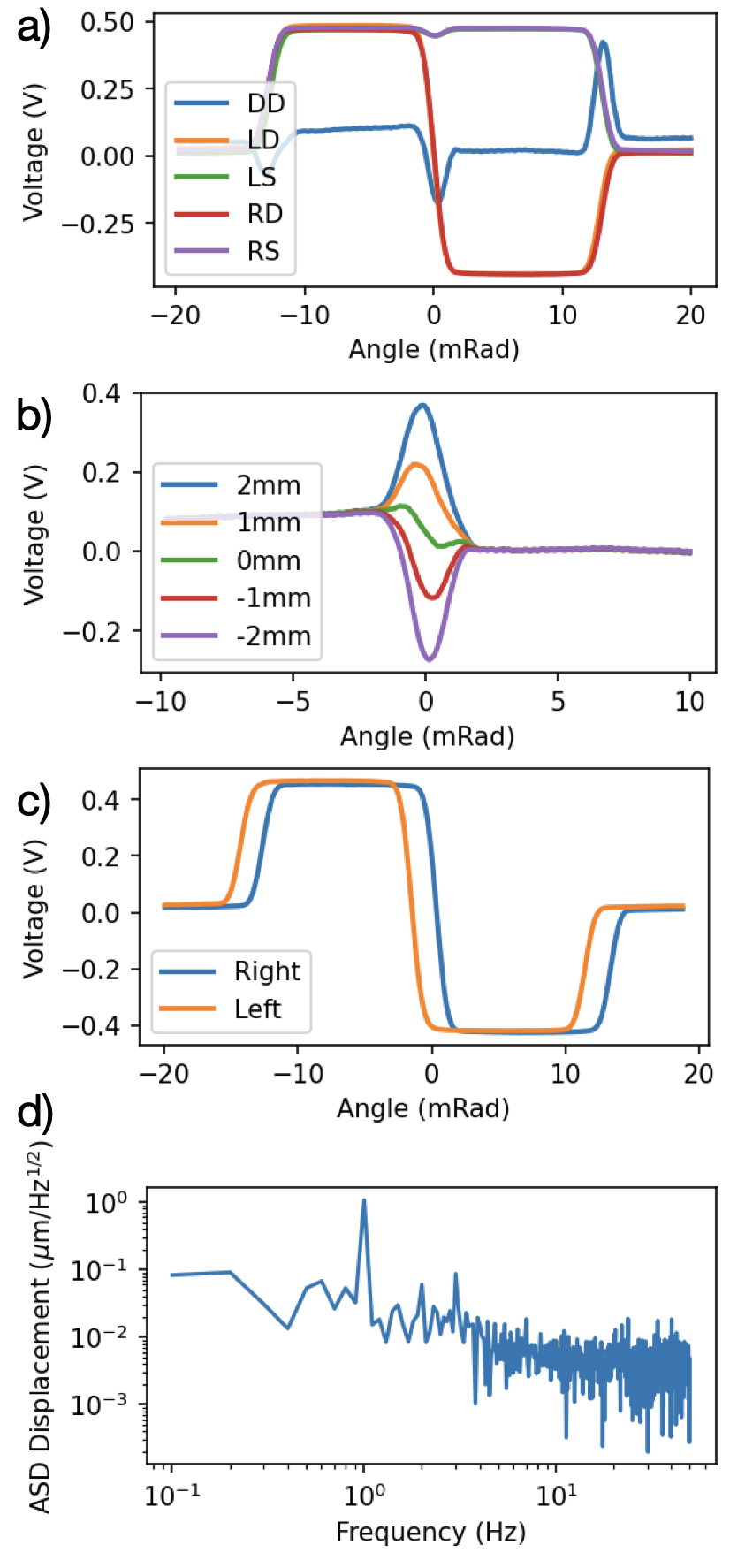}
    \caption{a) Sum and difference curves for the two balanced detectors as the platform rotates through the direction of the point source.  DD, LD, LS, RD, RS stand for double differential (6x amplification), left differential, left sum, right differential and right sum respectively. b) shows the angular scan for the double differential when the target is moved in increments of 1 mm from the center position. c) The LD and RD as the distance exceeds 4 cm from 1.42 m calibration point. d) Amplitude spectral density for a 1.1 micron amplitude sinusoidal oscillation about the 1.42 m mark based on a measured signal using 107 averages.}
    \label{fig:Results}
\end{figure}

Consider the measured results of the experiment shown in Fig. \ref{fig:Results}.  Fig. \ref{fig:Results} a)-c) were taken while the platform rotates through an angle that sweeps past the focal spot of the LED in the balanced detectors.  We followed a procedure for acquiring the scans.  The camera on the platform identifies the point source, the platform then moves to center the source on the detector plane and then the balanced detector-computer system does a fine-grained search for the precise direction. With the system centered, the platform is then angularly swept through small angular increments centerd on the source and the signals are acquired. Fig. \ref{fig:Results} a) shows the various measured signals of the system as the platform rotates through approximately 40 mRadians centered on the source. 

DD, RD, LD, RS, LS stand for double differential, right differential, left differential, right sum and left sum, respectively.  RD and LD are the differential signal coming from each balanced detector.  As perhaps expected from simultaneously zeroing both detectors on the source, the balanced signals, RD and LD, are almost identical (overlapping in the graph) modulo a small DC bias. Importantly, there is a steep voltage drop as the signal passes from one detector to the other detector.  The steepness of that slope and the noise characteristics determine the angular precision and ultimately depth precision.  

The double differential, DD, is the difference in the balanced signals coming from the two detectors.  The voltage differential was amplified by a factor of 6 and is shown in Fig. \ref{fig:Results} b) for several different distances relative to the 1.42 m calibration point. In this configuration, the angular discrepancies are based on $\beta_1$ and $\beta_2$. A Linear stage with 1 micron step sizes translated the source relative to the balanced detector platform in 1 mm increments.  At each increment the rotation stage scanned through a 20 mRad angular sweep.  

For the regime shown in Fig. \ref{fig:Results} b), the steep linear region of both balanced detectors overlap. As the source moves around the calibration point, the voltage differential strongly changes. A key takeaway from this Fig. \ref{fig:Results} is the relatively weak curvature of the curves centered around 0 mRad.  Cameras can easily give sub 100 microradian accuracy. The implication is that pointing a camera at a target, and using this double-differential regime, can result in a high accuracy single-shot distance estimate using a low-angular-accuracy camera pointing system. 

Fig. \ref{fig:Results} c) shows the left and right balanced signals when the steep slopes no longer overlap.  In this case, the source was moved 4cm from the calibration distance. For this scenario, one can still gain a high-accuracy distance measurement, but it requires knowing the relative separation of the curves.  For example, for unbiased detectors, one can simply measure the angular distance between the zero crossing of the two balanced signals.  Unfortunately, this again relies on the accuracy of the platform rotation to determine the depth and requires the platform to sweep through the point source.   

In Fig. \ref{fig:Results} d), the amplitude spectral density is measured for the source undergoing 1.1 micron RMS 1 Hz sinusoidal distance oscillations (the largest peak in the the spectrum).  These results were taken with the in-house detectors having an 11 cm baseline.  The small amplitude was achieved using a piezoactuated mount oscillating at 1 Hz (the peak in the spectrum).  The ASD was taken using a 10-second double-differential scan centered near the calibration point, averaged 107 times. The predicted shot noise limit for this system is about 50 pm$\text{Hz}^{-1/2}$.    So, while the measured precision is much better than existing systems, there is still room for improvement. 

\section{Discussion} \label{Discussion}

We have outlined a method for achieving remarkable passive depth estimation precision using a monolithic balanced detector system.  While the dual balanced detector system shows promise for single-point depth estimation, it should be clear that this system, in its current state, is a poor method for 3D mapping of complex targets. The precision of the system relies on analog differencing of the detectors prior to digitization. The implication of this work is that cameras that count photons at the shot noise limit have the potential to achieve fundamental triangulation limits for complex distributed targets albeit necessitating measurements with low light levels.  

While the main thrust of this work was a proof-of-principle demonstration of high precision triangulation, we envision a high accuracy system as well.  With careful attention to system calibration of the yaw angles and the amplification process, it is possible to not only be precise, but accurate. Consider a calibration which zeros the yaw angles ( $\alpha_1=\alpha_2=0$) by using two laterally-fixed-distance sources separated by the ideal baseline distance (one for each detector).  The detector positions and angles would be calibrated by moving the the dual source to two depths, and zeroed at both distances, sufficiently far apart to achieve the desired accuracy. Then, post-calibration, as long as a source is sufficiently distant, all distance measurements would be in this high-accuracy doubly-differential regime.  If we assume that the individual detectors have 1 mRad of linear voltage behavior as shown in Fig. \ref{fig:Results} a), b) and c), the object of interest must be at a distance greater than 100 meters from the detectors. Otherwise, the yaw angles must be nonzero, but calibrated, or the system must be scanned similar to that done in Fig. \ref{fig:Results} c).  We can shorten the minimum standoff distance by having the yaw angles each tilted towards each other and working around a calibration point or by slightly blurring the focal spot of the lens on the detector, but at the cost of precision. 

High accuracy, not just high precision, depth estimation over a long range leads us to another important demand on the system processing-- dynamic measurement range.  The current system uses a 24 bit ADC (1:16.7$\times 10^6$), but even having 1 nm resolution over a 1 meter distance is 9 orders of magnitude of dynamic range.  This demands an extremely low measurement noise floor combined with a high dynamic range ADC (e.g, 32 bits). This is the limiting factor for the dynamic range of our system.  

The experiment was performed on a non-floating optical table.  While significant effort was made to mitigate detection noise, no additional efforts were made to mitigate acoustic noise, turbulence, or temperature fluctuations.   It is difficult to make a noise budget for each of the noise sources.  However, in the ASD plot, we believe that above a few Hz, the floor is dominated by measurement noise. In the field, such noise sources will play a much bigger role in pointing precision and angle of arrival estimations. Further study is warranted in that domain.  The work does imply its potential as a satellite-based system where turbulence is negligible, although heating and cooling cycles would test calibration standards.  However, we believe this could be ameliorated as well with another onboard system calibration utilizing the same physics.

Lastly, the experimental emphasis has been on implementing monolithic design described in Fig. \ref{fig:Schematic} c).  We point out that the use of two systems as described in Fig. \ref{fig:Schematic} b) with nanoradian accuracy optical encoders or guidestars combined with a long baseline can achieve good accuracy over large distances even with moderate noise.  

\section{Conclusions} \label{Conclusions}

We have derived Cram\'er-Rao theoretical bounds for passive triangulation of a single point source using a lens detector system.  We showed that standard balanced detectors, owing to their analog differencing properties, can come close to this theoretical bound representing an improvement of several orders of magnitude over state-of-the-art depth detectors.  We then reported on a monolithic camera/balanced detector system with heretofore unrealized depth precision even for a relatively small baseline with respect to object distance.  We believe this can find use in many applications where targeting and tracking of a single point source is important.

\section{Acknowledgments}
JCH and ANJ acknowledge support from Chapman University.  

\section{Appendix: Three Dimensional Analysis}
\begin{figure}
    \centering
    \includegraphics[width=0.5\linewidth]{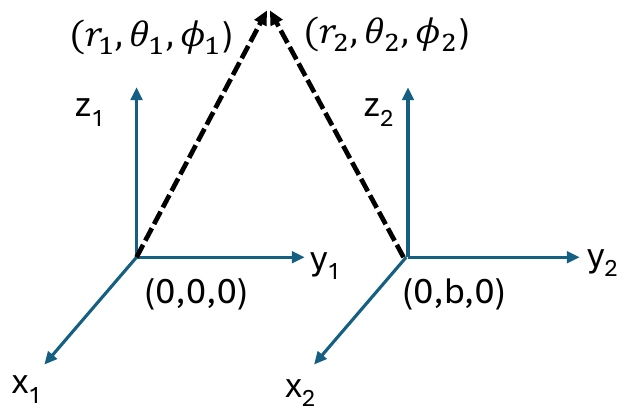}
    \caption{Defined coordinate system and detector locations.}
    \label{fig:coordinates}
\end{figure}

The analysis in the first section can be extended to three dimensions, which is the natural implementation of this problem.  
We position the first detector at the origin, and the second is displaced by a distance $b$ in the direction we define as $y$, and is at location $(0, b, 0)$, where we first use a Cartesian system.  The point source that is being imaged is located at coordinates $(x_1, y_2, z_2)$ for the first detector, and $(x_2, y_2, z_2)$ for the second detector.  We define the usual spherical polar coordinates $(r, \theta, \phi)$, where $\theta$ is the polar angle and $\phi$ is the azimuthal angle (physics convention), see Fig.~\ref{fig:coordinates}.  We use the subscript 1(2) referring to the spherical polar coordinates from the perspective of detector 1(2).

We will now calculate the differential change of the radius and angles of each detector relative to each other if the object being imaged is moved.  We first relate the coordinate systems to each other via their relative position as a translation, $(x_2, y_2, z_2) = (x_1, y_1 - b, z_1)$.  The relationship in spherical polar coordinates is more complicated.  After involved but straightforward trigonometric calculations, we find the relationships
\begin{eqnarray}
r_2 &=& \sqrt{ r_1^2 - 2 b r_1 \sin \theta_1 \sin \phi_1 + b^2}, \\
\tan \phi_2 &=& \tan \phi_1  - \frac{b}{r_1 \sin \theta_1 \cos \phi_1}, \\
\tan \theta_2  &=& \tan \theta_1 \sqrt{1 - \frac{2 b \sin \phi_1} {r_1 \sin \theta_1} + \frac{b^2}{r_1^2 \sin^2 \theta_1}}.  \label{convert}
\end{eqnarray}

To relate the differential changes, we note that $d {\bf r}_1 = d{\bf r_2}$ in Cartesian coordinates.  We can relate the changes in Cartesial coordinate to changes in the spherical polar coordinates via the Jacobian matrix, $(dx, dx, dz)^T = J \cdot (dr, d\theta, d\phi)^T$.  The expression for the Jacobian matrix in spherical polar coordinates is
\begin{equation}
J = \frac{d(x, y, z)}{d(r, \theta, \phi)} = \begin{pmatrix} \sin \theta \cos \phi, & r \cos \theta \cos \phi, & - r \sin \theta \sin \phi \\ \sin \theta \sin \phi, & r \cos \theta \sin \phi, &  r \sin \theta \cos \phi \\ 
\cos \theta, & -r \sin \theta, & 0 \end{pmatrix} \nonumber
\end{equation}

We can then relate the changes of coordinate system 1 to coordinate system 2 in the spherical polar coordinates as
\begin{equation}
J_1 \cdot \begin{pmatrix} d r_1 \\ d \theta_1 \\ d \phi_1 \end{pmatrix}=J_2 \cdot \begin{pmatrix} d r_2 \\ d \theta_2 \\ d \phi_2 \end{pmatrix},
\end{equation}
where the index represents writing the Jacobian matrix with coordinate labels 1 or 2.
We can then relate the changes from the second system in terms of the first via the matrix inverse,
\begin{equation}
\begin{pmatrix} d r_2 \\ d \theta_2 \\ d \phi_2 \end{pmatrix} = J_2^{-1} \cdot J_1 \cdot \begin{pmatrix} d r_1 \\ d \theta_1 \\ d \phi_1 \end{pmatrix}.
\end{equation}

Using cofactor matrix methods, the matrix inverse of the Jacobian matrix is
\begin{equation}
J^{-1}  = 
    \begin{pmatrix} \sin \theta \cos \phi, & \sin \theta \sin \phi, &  \cos \theta  \\ \frac{1}{r} \cos \theta \cos \phi, & \frac{1}{r} \cos \theta \sin \phi, &  -\frac{1}{r} \sin \theta \\ 
-\frac{\sin \phi}{r \sin \theta}, & \frac{\cos \phi}{r \sin \theta}, & 0 \end{pmatrix}. \nonumber
\end{equation}
This translation completes the formal and general solution of the problem.

It is of interest to consider the case when $b$ is small, to get the leading order behavior.  We express elements of the conversion matrix in terms of only the first detector's coordinates by the translation (\ref{convert}),
\begin{eqnarray}
(J_2^{-1} \cdot J_1)_{11} &=& 
1 - (3 + \cos(2 \phi_1) + 2 \cos(2 \theta_1) \sin^2 \phi_1))\frac{b^2}{8 r_1^2} \nonumber\\
(J_2^{-1} \cdot J_1)_{12} &=& -\cos(\theta_1) \sin(\phi_1) b \nonumber \\
(J_2^{-1} \cdot J_1)_{13} &=& -\sin(\theta_1) \cos(\phi_1)
 b \nonumber \\
(J_2^{-1} \cdot J_1)_{21} &=& \cos(\theta_1) \sin(\phi_1)\frac{ b}{r_1^2} \nonumber \\
(J_2^{-1} \cdot J_1)_{22} &=& 1 + \sin(\theta_1) \sin(\phi_1)\frac{ b}{r_1} \nonumber \\
(J_2^{-1} \cdot J_1)_{23} &=&-\cos(\theta_1) \cos(\phi_1)\frac{ b}{r_1} \nonumber \\
(J_2^{-1} \cdot J_1)_{31} &=&  \csc(\theta_1) \cos(\phi_1)\frac{ b}{r_1^2} 
\nonumber \\
(J_2^{-1} \cdot J_1)_{32} &=& \csc(\theta_1) \cot(\theta_1) \cos(\phi_1)\frac{ b}{r_1} \nonumber \\
(J_2^{-1} \cdot J_1)_{33} &=& 1 + \csc(\theta_1) \csc(\theta_1) \sin(\phi_1)\frac{ b}{r_1}\nonumber \\
\end{eqnarray}

\vfill{\eject}
\bibliography{Main}

\end{document}